\documentclass[aps,prb,twocolumn]{revtex4}

\usepackage{graphicx}%
\usepackage{dcolumn}
\usepackage{amsmath}
\usepackage{bm}
\makeatletter
\def\btt#1{\texttt{\@backslashchar#1}}%
\DeclareRobustCommand\bblash{\btt{\@backslashchar}}%
\makeatother
\begin{document}

\title[Short Title]{Generic gauge fields  in  the Hubbard model:
emergence of pairing interaction}
 
 \date{\today}

\begin{abstract}
The  spin-rotationally invariant SU(2) approach  to the Hubbard model is extended
to accommodate  the charge degrees of freedom.
Both U(1) and SU(2) gauge transformation are useed
to  factorize the charge and spin contribution to the original electron operator
in terms of the  emergent gauge fields. It is shown that these fields play a similar role
as  phonons in the BCS theory:  they provide the ``glue" for fermion pairing.
By tracing out gauge bosons the form of paired states is established and the role
 of antiferromagnetic correlations is explicated.

\end{abstract}

\author{T. K. Kope\'{c}}
\affiliation{
Institute for Low Temperature and Structure Research,Polish Academy of Sciences,\\
POB 1410, 50-950 Wroclaw 2, Poland}
\maketitle	
To understand the physics
of strongly correlated  (SC) systems one frequently employs a "slave
particle" (SP) decomposition of the electron operator.\cite{sp1,sp2}
 Analytic theory behind this formulation hinges on the treatment of the constraint of
no double occupation based on the assumption that the on-site interaction energy $U$ can be
renormalized to infinity. The redundancy in  representations
used to enforce the constraint naturally leads to
various gauge theories.\cite{gauge}
It is sometimes supposed that slave particles can be
liberated at low energies,  in which case
the slave-boson and fermion degrees of freedom  take on a physical
meaning with the spin-charge separation as a result.
 However,  it was shown that   gauge
theories associated with SP representations of
correlated electrons, such as the $t-J$ model, are always
confining.\cite{nayak} The reason is that
the slave-particle
gauge theory is infinitely strongly coupled-- there is no
intrinsic kinetic energy for the gauge field.
Therefore, the slave particle representations of correlated
electron models are not belonging to the class of {\it generic }	gauge theories.
Although matter fields which couple to the non-generic SP gauge field might generate 
gauge dynamics in the low-energy effective action,\cite{ichinose}  the problem remains:
how to {\it derive}  emergent gauge fields
from the microscopic formulation of the theory of SC
electrons and  not just to  write down a Lagrangian
that contains them  - as in the SP approach. 
This issue is of basic concern because 
the understanding of the mechanism of
superconductivity in  cuprates requires the knowledge  of  bosons mediating
the pairing as well as 	the nature of the paired states.
Here, the  underlying attraction force appears very puzzling since it is hard to reconcile
the  microscopic attractive interaction  with the completely repulsive bare
electron-electron forces.

In the present paper we extend the  SU(2) spin-rotationally invariant approach 
to the Hubbard model,\cite{schulz} which makes no  assumptions regarding the magnitude of the Coulomb energy $U$,  to accommodate on equal footing the {\it charge} degrees of freedom. Using U(1) and SU(2) transformation we explicitly factorize the charge and spin contribution to the original electron operator in terms of the corresponding {\it emergent} gauge fields. We show that these fields play a similar role as  phonons in the BCS theory:  they provide the ``glue" for fermion pairing in the SC system. By tracing out gauge bosons we  explicitly calculate and  the
form of paired states and explicate the role of antiferromagnetic (AF) correlations.

Our starting point is the purely fermionic Hubbard Hamiltonian
 ${\cal H}\equiv {\cal H}_t+{\cal H}_U$:
\begin{eqnarray}
{\cal H}=
 -t\sum_{\langle {\bf r}{\bf r}'\rangle,\alpha}
 [c^{\dagger }_{{\alpha}}({\bf r})
c_{\alpha }({\bf r}')+{\rm h.c.}]
+
\sum_{\bf r}Un_{\uparrow} ({\bf r}) n_{\downarrow}({\bf r}).
\label{mainham}
\end{eqnarray}
Here, $\langle {\bf r},{\bf r}'\rangle$  runs
over the nearest-neighbor (n.n.) sites, $t$  is the  hopping amplitude, $U$ stands for
the Coulomb repulsion,
while the operator $c_{\alpha }^\dagger({\bf r})$
creates an electron with spin $\alpha=\uparrow,\downarrow$ at the lattice site ${\bf r}$.
Furthermore,
${n}({{\bf r}})= n_{\uparrow} ({\bf r})+n_{\downarrow}({\bf r})$
is the  number operator, where  ${n}_{\alpha}({{\bf r}})= c^\dagger_{\alpha}({\bf r})
c_{\alpha}({\bf r})$. Usually, working in the grand canonical ensemble  a term
$-\mu\sum_{\bf r}{n}({\bf r})$ is added to ${\cal H}$ in Eq.(\ref{mainham})
with $\mu$  being  the chemical potential .
It is customary to introduce Grassmann fields,
$c_\alpha({\bf r}\tau)$ 
depending on the ``imaginary time" $0\le\tau\le \beta\equiv 1/k_BT$,
(with $T$ being the temperature)
that satisfy the anti--periodic condition
$c_{\alpha}({\bf r}\tau)=-c_{\alpha}({\bf r}\tau+\beta)$,
to write the path integral for the statistical sum
${\cal Z}=\int\left[{\cal D}\bar{c}  {\cal D}{c}
\right]e^{-{\cal S}[\bar{c},c]}$
with the fermionic action 
\begin{eqnarray}
{\cal S}[\bar{c},c]={\cal S}_B[\bar{c},c]+\int_0^\beta d\tau{\cal H}[\bar{c},c],
 \end{eqnarray}
that contains the fermionic  Berry term
\begin{eqnarray}
{\cal S}_B[\bar{c},c]=\sum_{{\bf r}\alpha }\int_0^\beta d\tau
 \bar{c}_{\alpha }({\bf r}\tau)\partial_\tau{c}_{\alpha }({\bf r}\tau)
\label{cberry}.
 \end{eqnarray}
For the SC system it is  crucial to construct a covariant formulation of the theory
which  naturally  preserves  the  spin-rotational symmetry present in the Hubbard Hamiltonian.
For this purpose the density--density product 
in Eq.(\ref{mainham}) we write, following Ref.\onlinecite{schulz}, in
a spin-rotational invariant way:
\begin{equation}
{\cal H}_U=U\sum_{{\bf r} }\left\{\frac{1}{4}{n }^2({{\bf r}}\tau)
-\left[{\bf \Omega} ({\bf r}\tau)\cdot{\bf S} ({\bf r}\tau)\right]^2\right\},
\label{huu}
\end{equation}
where
 $S^a({\bf r}\tau)=\frac{1}{2}\sum_{\alpha\alpha'}c^\dagger_{\alpha}({\bf r}\tau)
\hat{\sigma}_{\alpha\alpha'}^a c_{\alpha'}({\bf r}\tau)$
denotes the vector spin operator ($a=x,y,z$) with $\hat{\sigma}^a$ being
the Pauli matrices. The unit vector 
 ${\bf \Omega} ({\bf r}\tau)=[\sin\vartheta({\bf r}\tau)\cos\varphi({\bf r}\tau),
\sin\vartheta({\bf r}\tau)\sin\varphi({\bf r}\tau),
\cos\vartheta({\bf r}\tau)]$
written in terms of polar angles labels
varying in space-time  spin quantization axis. 
The spin--rotation invariance is made explicit
by performing the angular integration over  ${\bf\Omega}({\bf r}\tau)$
at each site and time. By decoupling spin and charge density terms 
in Eq.(\ref{huu}) using auxiliary fields
$\varrho({\bf r}\tau)$ and $iV({\bf r}\tau)$ respectively,
we write down the partition function in the form
\begin{eqnarray}
{\cal Z}&=&\int[{\cal D}{\bf \Omega}]\int[{\cal D} V{\cal D}{\varrho}
 ]\int\left[
  {\cal D}\bar{c}{\cal D}c
\right]\times
\nonumber\\
&\times&
e^{-{\cal S}\left[{\bf \Omega},V,{\varrho},\bar{c},c\right]}.
\label{zfun}
\end{eqnarray}
where$[{\cal D}{\bf \Omega}]\equiv
\prod_{{\bf r}\tau_k}
\frac{\sin\vartheta({\bf r}\tau_k)d\vartheta ({\bf r}\tau_k)d\varphi ({\bf r}\tau_k)}{4\pi}$
is the spin-angular integration measure.
The  effective action reads:
\begin{eqnarray}
{\cal S}\left[{\bf \Omega},V,{\varrho},\bar{c},c\right]&=&
\sum_{ {\bf r} }\int_0^\beta
 d\tau
\left[\frac{{\varrho}^2 ({\bf r}\tau)}{U}+\frac{V^2 ({\bf r}\tau)}{U}\right.
\nonumber\\
&+&\left.iV ({\bf r}\tau)n ({\bf r}\tau)
+2{\varrho} ({\bf r}\tau){\bf \Omega} ({\bf r}\tau)\cdot {\bf S} ({\bf r}\tau)\right]
\nonumber\\
&+&{\cal S}_B[\bar{c},c]+\int_0^\beta d\tau{\cal H}_t[\bar{c},c].
\label{sa}
\end{eqnarray}
Simple Hartree-Fock (HF) theory will not work for a Hubbard model in which $U$
 is the largest energy in the problem.
One has to isolate strongly fluctuating modes generated by the Hubbard term  according to the
charge U(1) and spin SU(2) symmetries.
To this end  we write the fluctuating ``imaginary chemical potential" $iV ({\bf r}\tau)$ as a sum of
a static $V_{0 }({\bf r})$ and periodic function
$V({\bf r}\tau)=V_0({\bf r})+\tilde{V}({\bf r}\tau)$
using  Fourier series
\begin{eqnarray}
\tilde{V}({\bf r}\tau)=\frac{1}{\beta}	\sum_{n=1}^\infty
[\tilde{V}({\bf r}\omega_n)e^{i\omega_n\tau}+c.c.]
\label{decomp}
\end{eqnarray}
with $\omega_n=2\pi n/\beta$ ($n=0,\pm1,\pm2$)
being the (Bose) Matsubara frequencies.
Now, we introduce the U(1) {\it phase } field ${\phi} ({\bf r}\tau)$
via the Faraday--type relation
\begin{equation}
\dot{\phi} ({\bf r}\tau)\equiv\frac{\partial\phi ({\bf r}\tau)}
{\partial\tau}=\tilde{V} ({\bf r}\tau).
\label{jos}
\end{equation}
Since the homotopy group $\pi_1[U(1)]$ forms a set of integers,
discrete configurations of $\phi({\bf r}\tau)$ matter, for which
$\phi({\bf r}\beta)-\phi({\bf r}0)=2\pi m({\bf r})$, where $m({\bf r})=0,\pm 1,\pm 2,\dots$
Thus the decomposition  of the charge field
$V({\bf r}\tau)$   conforms with the  basic $m=0$ topological sector since
$\int_0^\beta\dot{\phi} ({\bf r}\tau)=\int_0^\beta \tilde{V} ({\bf r}\tau)\equiv 0$.
Furthermore, by performing the local gauge transformation to the {\it new} fermionic
variables $f_{\alpha}({\bf r}\tau)$:
\begin{eqnarray}
\left[\begin{array}{c}
c_{\alpha }({\bf r}\tau)\\
\bar{c}_{\alpha }({\bf r}\tau)
\end{array}\right]=
\left[\begin{array}{cc}
z({\bf r}\tau)&0\\
0& \bar{z}({\bf r}\tau)
\end{array}\right]
\left[\begin{array}{c}
f_{\alpha }({\bf r}\tau)\\
\bar{f}_{\alpha }({\bf r}\tau)
\end{array}\right]
\label{sing1}
\end{eqnarray}
where the unimodular parameter  $|z({\bf r}\tau)|^2=1$ satisfies $z({\bf r}\tau)=e^{i\phi ({\bf r}\tau)}$,
we remove the imaginary term $i\int_0^\beta d\tau\tilde{V}({\bf r}\tau)n({\bf r}\tau)$
for all the Fourier modes
of the $V ({\bf r}\tau)$ field, except for  the zero frequency.
Subsequent SU(2) transformation from $f_{\alpha}({\bf r}\tau)$ to  $h_{\alpha}({\bf r}\tau)$
operators,
\begin{eqnarray}
\left[\begin{array}{c}
f_{1 }({\bf r}\tau)\\
{f}_{2}({\bf r}\tau)
\end{array}\right]=
\left[
\begin{array}{cc}
\zeta_{ 1}({\bf r}\tau) & -\bar{\zeta}_{2}({\bf r}\tau) \\
\zeta_{2}({\bf r}\tau) & \bar{\zeta}_{1}({\bf r}\tau)
\end{array}
\right]\left[\begin{array}{c}
h_{1 }({\bf r}\tau)\\
{h}_{2}({\bf r}\tau)
\end{array}\right]
\label{sing2}
\end{eqnarray}
with the constraint
$|\zeta_{1}({\bf r}\tau)|^2 +|\zeta_{2}({\bf r}\tau)|^2=1$ takes away the
rotational dependence on ${\bf \Omega}({\bf r}\tau)$ in the spin sector.
This is done
by means  of the Hopf map
${\bf R}({\bf r}\tau) \hat{\sigma}^z{\bf R}^\dagger({\bf r}\tau) =\hat{{\bm\sigma}}\cdot{\bf \Omega}({\bf r}\tau)$
that is based on the enlargement from two-sphere $S_2$ to the three-sphere $S_3\sim {\rm SU}(2)$.
The unimodular constraint
can be resolved by using the parametrization
\begin{eqnarray}
\zeta_{1}({\bf r}\tau)& = & e^{-\frac{i}{2}[\varphi({\bf r}\tau)+\chi({\bf r}\tau)]}
\cos\left[\frac{\vartheta({\bf r}\tau)}{2}\right]
\nonumber\\
\zeta_{2}({\bf r}\tau)&=&e^{\frac{i}{2}[\varphi({\bf r}\tau)-\chi({\bf r}\tau)]}
\sin\left[\frac{\vartheta({\bf r}\tau)}{2}\right]
\label{cp1}
\end{eqnarray}
with the  Euler
angular variables $\varphi({\bf r}\tau),\vartheta({\bf r}\tau)$ and $\chi({\bf r}\tau)$, respectively.
Here, the  extra variable $\chi({\bf r}\tau)$   represents the U(1) gauge freedom  of the theory
as a consequence of $S_2\to S_3$ mapping. One can summarize Eqs (\ref{sing1}) and (\ref{sing2})
by the single joint gauge transformation exhibiting electron operator factorization
\begin{eqnarray}
c_{\alpha }({\bf r}\tau)=\sum_{\alpha'}z({\bf r}\tau)
{ R}_{\alpha\alpha'} ({\bf r}\tau)h_{\alpha'}({\bf r}\tau),
\label{decomp2}
\end{eqnarray}
where ${\bf R}({\bf r}\tau) =e^{-i\hat{\sigma}_z\varphi({\bf r}\tau)/2}e^{-i\hat{\sigma}_y\vartheta({\bf r}\tau)/2}
e^{-i\hat{\sigma}_z\chi({\bf r}\tau)/2}$
is  a unitary matrix which
rotates the spin-quantization axis at site ${\bf r}$ and time $\tau$.
Eq.(\ref{decomp2}) reflects the composite nature of the interacting electron formed from
bosonic spinorial and charge degrees of freedom given by  ${ R}_{\alpha\alpha'} ({\bf r}\tau)$
and $z({\bf r}\tau)$, respectively as well as  remaining fermionic part $h_{\alpha}({\bf r}\tau)$.
In the new variables the action in Eq.(\ref{sa}) assumes the form
\begin{eqnarray}
&&{\cal S}\left[{\bf \Omega},\phi,{\varrho},\bar{h},h\right]=
{\cal S}_B[\bar{h},h]+\int_0^\beta d\tau{\cal H}_{\bf \Omega,\phi}[\rho,\bar{h},h] 
\nonumber\\
&&+
{\cal S}_0\left[\phi\right]+2\sum_{\bf r }\int_0^\beta d\tau
{\bf A}({\bf r}\tau)\cdot {\bf S}_{h }({\bf r}\tau),
\label{sa2}
\end{eqnarray}
where ${\bf S}_{h}({\bf r}\tau)=\frac{1}{2}\sum_{\alpha\gamma}\bar{h}_{\alpha }({\bf r}\tau)
\hat{\bm\sigma}_{\alpha\gamma} h_{\gamma}({\bf r}\tau)$. Furthermore,
\begin{eqnarray}
S_0[\phi]=\sum_{\bf r}\int_0^\beta d\tau\left[\frac{\dot{\phi}^2({\bf r}\tau)}{U} 
+\frac{2\mu}{iU}\dot{\phi}({\bf r}\tau) \right]
\label{sphi}
\end{eqnarray}
stands for the kinetic  and Berry term of the U(1) phase field in the charge sector.
The SU(2) gauge transformation in Eq.(\ref{sing2}) and the  fermionic Berry term in Eq.(\ref{cberry})
generate SU(2) potentials given  by
${\bf R} ^\dagger({\bf r}\tau){\partial}_{\tau}
{\bf R} ({\bf r}\tau) =-
{\hat{\bm\sigma}}\cdot {\bf A}({\bf r}\tau)$,
where
\begin{eqnarray}
A^x({\bf r}\tau)&=&\frac{i}{2}\dot{\vartheta}({\bf r}\tau)\sin\chi({\bf r}\tau)
\nonumber\\
&-&\frac{i}{2}\dot{\varphi}({\bf r}\tau)\sin\theta({\bf r}\tau)\cos\chi({\bf r}\tau)
\nonumber\\
A^y({\bf r}\tau)&=&\frac{i}{2}\dot{\vartheta}({\bf r}\tau)\cos\chi({\bf r}\tau)
\nonumber\\
&+&\frac{i}{2}\dot{\varphi}({\bf r}\tau)\sin\theta({\bf r}\tau)\sin\chi({\bf r}\tau)		
\nonumber\\
A^z({\bf r}\tau)&=&\frac{i}{2}\dot{\varphi}({\bf r}\tau)\cos\vartheta({\bf r}\tau)
+\frac{i}{2}\dot{\chi}({\bf r}\tau).
\end{eqnarray}
In analogy to the charge U(1) field the SU(2)  spin system exhibits  emergent dynamics.
By integrating out fermions the last term in Eq.(\ref{sa2}) will generate the kinetic term
for the SU(2) rotors ${\cal S}_0[{\bf \Omega}]=-({1}/{{\cal E}_s})\sum_{\bf r}
\int_0^\beta  d \tau {\bf A}({\bf r}\tau)\cdot{\bf A}({\bf r}\tau)$ in the form
\begin{eqnarray}
&&{\cal S}_0[{\bf \Omega}]=\frac{1}{{4\cal E}_s}\sum_{\bf r}\int_0^\beta  d \tau 
\left[ \dot{\vartheta}^2({\bf r}\tau)+\dot{\varphi}^2({\bf r}\tau)
\right.
\nonumber\\
&&+
\left.\dot{\chi}^2({\bf r}\tau) +2\dot{\varphi}({\bf r}\tau)
\dot{\chi}({\bf r}\tau)\cos{\vartheta}({\bf r}\tau)
\right]
\label{phiaction}
\end{eqnarray}
with ${\cal E}_s$ being  of order of $U$ close to half filling.
The  first order term in ${\bf A}({\bf r}\tau)$ fields gives rise to the
usual spin Berry contribution. If we  work in  Dirac ``north pole" gauge 
${\chi}({\bf r}\tau)=-{\varphi}({\bf r}\tau)$ 
one recoverss the familiar form
${\cal S}_{B}[{\bf \Omega}]=-{i({\rho}/{U}})\sum_{\bf r}\int_0^\beta  d \tau
\dot{\varphi}({\bf r}\tau)[1-\cos\vartheta({\bf r}\tau)]$.
The fermionic sector, in turn, is governed by the effective Hamiltonian
\begin{eqnarray}
 &&{\cal H}_{\bf \Omega,\phi} = 
	\sum_{{ \bf r}}{\varrho} ({\bf r}\tau) [\bar{h}_{{\uparrow} }({\bf r}\tau)h_{\uparrow  }({\bf r}\tau)-
\bar{h}_{{\downarrow} }({\bf r}\tau)h_{\downarrow  }({\bf r}\tau)]
\nonumber\\
&&-t\sum_{\langle {\bf r},{\bf r}'\rangle\atop \alpha\gamma} \bar{z}({\bf r}\tau)z({\bf r}'\tau)
 \left[{\bf R}^\dagger ({\bf r}\tau){\bf R}_{ }({\bf r'}\tau)\right]_{\alpha\gamma}
\bar{h}_{{\alpha} }({\bf r}\tau)h_{\gamma }({\bf r}'\tau)
\nonumber\\
&&-\bar{\mu}\sum_{{ \bf r}\alpha}
\bar{h}_{\alpha }({\bf r}\tau)
 h_{\alpha }({\bf r}\tau),
\label{explicit}
\end{eqnarray}
where $\bar{\mu}=\mu-n_fU/2$ is the chemical potential
with a Hartree shift originating from the saddle-point value of the static variable $V_0({\bf r})$ with $n_h=n_{h\uparrow}+n_{h\downarrow}$ and $n_{h\alpha}=
\langle\bar{h}_{\alpha }({\bf r}\tau)h_{\alpha }({\bf r}\tau)\rangle$.
The chief merit of the gauge transformation
in Eq.(\ref{decomp2}) is that we have managed to cast the SC
problem into a system of non-interacting $h$ fermions
submerged in the bath of strongly fluctuating U(1) and SU(2) gauge potentials coupled to 
fermions via hopping term plus Zeeman-type contribution with the massive field
${\varrho} ({\bf r}\tau)$. In the AF phase, at the half-filling, it assumes the
staggered form ${\varrho} ({\bf r}\tau)=\Delta_c (-1)^{\bf r}$ with $\Delta_c$ being the
charge gap $\Delta_c\sim U/2$ for $U/t\gg 1$.
However, a nonzero  value of $\Delta_c$ does not  imply the existence of AF  long--range 
order. For this the angular degrees of freedom ${\bf \Omega}({\bf r}\tau)$ have also to be ordered,
whose  low-lying  excitations  are in the form of spin waves.

It is well known that phonons
play the role of the ``glue" that is responsible for the formation of 
Cooper pairs in  conventional  superconductors. Now we show that U(1) {\it and} SU(2)
emergent gauge fields, the collective high energy modes in the SC system,  take
over the task which was carried out by phonons in BCS superconductors.
In a way similar to phonons these  gauge fields  couple to the fermion  density  type term 
via the   amplitude $t$, see Eq.(\ref{explicit}).
Now we evaluate  the effective interaction between fermions by
tracing out the gauge degrees of freedom.
To this end we write the partition function as ${\cal Z}=\int[{\cal D}{\bar h}{\cal D}h]e^{-{\cal S}[\bar{h},h]}$,
where
\begin{eqnarray}
{\cal S}[\bar{h},h]&=&-\ln
\int\left[ {\cal D}{\bf \Omega}{\cal D}\phi\right]e^{-{\cal S}[{\bf \Omega},\phi,\bar{h},h]}
\label{x1}
\end{eqnarray}
generates cumulant expansion  for the effective fermionic action. We concentrate on
the second order term in the hoping amplitude $t$ containing four fermion operators:
\begin{widetext}
\begin{eqnarray}
 {\cal S}^{(2)}[{\bar h},h] =&&-\frac{t^2}{2}
\int_0^\beta d\tau d\tau'\left\langle 
\sum_{ |{\bf r}_1-{\bf r}_1'|=n.n.}\bar{z}({\bf r}_1\tau)z({\bf r}_1'\tau)
 \sum_{\alpha \alpha'}
 \left[{\bf R}^\dagger({\bf r}_1\tau){\bf R}_{ }({\bf r'}_1\tau)\right]_{\alpha\alpha'}
\bar{h}_{{\alpha} }({\bf r}_1\tau)h_{\alpha'  }({\bf r}_1'\tau)\right.\times
\nonumber\\
&&\times\left. \sum_{ |{\bf r}_2-{\bf r}_2'|=n.n.}\bar{z}({\bf r}_2\tau')z({\bf r}_2'\tau')
 \sum_{\gamma\gamma'}
 \left[{\bf R}^\dagger ({\bf r}_2\tau'){\bf R}({\bf r}_2'\tau')\right]_{\gamma\gamma'}
\bar{h}_{{\gamma} }({\bf r}_2\tau')h_{\gamma'  }({\bf r}_2'\tau')\right\rangle,
\end{eqnarray}
\end{widetext}
where $\langle\dots\rangle$ denotes averaging over U(1) and SU(2) gauge fields.
The averaging  in the charge sector is performed with the
use of the  U(1) phase action in Eq.(\ref{sphi}) to give
\begin{eqnarray}
&&\langle\bar{z}({\bf r}_1\tau)z({\bf r}_1'\tau)\bar{z}({\bf r}_2\tau')z({\bf r}_2'\tau') \rangle
\nonumber\\
&&\simeq(\delta_{\bf r_1,r_1'}\delta_{\bf r_2,r_2'}+\delta_{\bf r_1,r_2'}
\delta_{\bf r'_1,r_2})\times
\nonumber\\
&&\times\exp\left\{-\frac{U}{2}\left[|\tau-\tau'|
	-\frac{(\tau-\tau')^2}{\beta} \right]  \right\}.
	\label{chargecharge}
\end{eqnarray}
Equation (\ref{chargecharge}) reflects the local (in space) nature of charge excitation
and contains only the non-topological part of the four-point charge correlator.
Away from half-filling the  dynamics of spin variables is slower 
as compared to the charge counterparts,
allowing to treat  SU(2) variables as local in time
${\bf R}({\bf r}\tau')={\bf R}_{ }({\bf r}\tau)+(\tau'-\tau)\partial_\tau
{\bf R}({\bf r}\tau) + O[(\tau'-\tau)^2]$.
Furthermore, in the low temperature limit (on the energy scale given by $U$),
by making use of the formula
\begin{eqnarray}
\lim_{\tau\to 0}\int_{0}^{\beta}d\tau'{e^ {-{ {{\left| \tau-\tau'\right| U}\over{2}} }}}={{2}\over{U}}-{{2e^ {-{ {{\beta U}\over{2}} }}}\over{U}} 
\end{eqnarray}
we arrive at
\begin{eqnarray}
&&{\cal S}^{(2)}[{\bar h},h] =
-\frac{t^2}{U}\int_0^\beta d\tau \sum_{\langle {\bf r},{\bf r}'\rangle}
 \sum_{\alpha\alpha'\atop \gamma\gamma'}
[\bar{h}_{{\alpha}}({\bf r}'\tau)h_{\alpha'}({\bf r}\tau) ]^\dagger 
\nonumber\\
&&\times\left\langle M_{\alpha'\alpha;\gamma\gamma'}({\bf r}\tau;{\bf r'}\tau  |{\bf r'}\tau{\bf r}\tau)
\right\rangle \bar{h}_{{\gamma}}({\bf r}'\tau)h_{\gamma '}({\bf r}\tau) ,
\end{eqnarray}
where $\langle\dots\rangle$ denotes averaging over the remaining  spin-angular variables
and
\begin{eqnarray}
M_{\alpha'\alpha;\gamma\gamma'}=\left[{\bf R}^\dagger({\bf r}\tau){\bf R}({\bf r'}\tau)\right]_{\alpha'\alpha}
 \left[{\bf R}^\dagger({\bf r}'\tau){\bf R}({\bf r}\tau)\right]_{\gamma\gamma'}.
\end{eqnarray}
Now, employing the composition formula for rotational matrices\cite{fradkin}
\begin{eqnarray}
{\bf R}^\dagger({\bf r}\tau){\bf R}({\bf r'}\tau) =\frac{1}{\sqrt{2}}\left[
\begin{array}{cc}
{ e^{\frac{i}{2}\Phi}}{\Upsilon_+}
 & { e^{\frac{i}{2}\bar{\Phi}}}{\Upsilon_-}
\\
-{e^{-\frac{i}{2}\bar{\Phi}}}{\Upsilon_-}
 & {e^{-\frac{i}{2}\Phi}}{\Upsilon_+}
\end{array}
\right]
\end{eqnarray}
with
$\Upsilon_\pm({\bf r}\tau,{\bf r'}\tau)=\sqrt{1\pm{\bf \Omega}
({\bf r}\tau)\cdot{\bf \Omega}({\bf r'}\tau)}$,
where $\Phi\equiv\Phi[{\bf \Omega}({\bf r}\tau),{\bf \Omega}({\bf r'}\tau ),{\bf z}]$
is the signed solid angle spanned by the vectors ${\bf \Omega}({\bf r}\tau),{\bf \Omega}({\bf r'}\tau )$ and {\bf z} with $\bar{\Phi}=\Phi[{\bf \Omega}({\bf r}\tau),-{\bf \Omega}({\bf r'}\tau )]
-2\varphi({\bf r}\tau)$, we finally conclude that
\begin{eqnarray}
{\cal S}^{(2)}[{\bar h},h] &=&
- \sum_{\langle {\bf r}{\bf r}'\rangle}\int_0^\beta d\tau 
 \left[ 
J_A\bar{\cal A}({\bf r}\tau{\bf r}'\tau){\cal A}({\bf r}\tau{\bf r}'\tau) \right.
 \nonumber\\
&+& \left.J_F\bar{\cal F}({\bf r}\tau{\bf r}'\tau){\cal F}({\bf r}\tau{\bf r}'\tau)\right],
\label{pairing}
\end{eqnarray}
where
\begin{eqnarray}
&&J_{A/F}=\frac{2t^2}{U}
\left[ 1\mp \langle{\bf\Omega}({\bf r}\tau)
\cdot{\bf \Omega}({\bf r'}\tau )  \rangle\right]
\label{jaf}
\end{eqnarray}
and
\begin{eqnarray}
&&{\cal A} ({\bf r}\tau{\bf r}'\tau)=\frac{h_{  \uparrow}({\bf r}\tau)h_{ \downarrow}({\bf r'}\tau)
-h_{  \downarrow}({\bf r}\tau) h_{  \uparrow}({\bf r'}\tau)}{\sqrt{2}}
\nonumber\\
&&{\cal F} ({\bf r}\tau{\bf r}'\tau)=
\frac{{\bar h}_{  \uparrow}({\bf r}\tau)h_{ \uparrow}({\bf r'}\tau)
+{\bar h}_{  \downarrow}({\bf r}\tau) h_{  \downarrow}({\bf r'}\tau)}{\sqrt{2}}
\end{eqnarray}
being the valence bond operators.\cite{rvb} The rotational invariance of the right-hand side in
Eq.(\ref{pairing}) is manifest since
$-\bar{\cal A}({\bf r}\tau{\bf r}'\tau){\cal A}({\bf r}\tau{\bf r}'\tau)=
{\bf S}_{h}({\bf r}\tau)\cdot{\bf S}_{h}({\bf r'}\tau )-\frac{1}{4}$ and
$\bar{\cal F}({\bf r}\tau{\bf r}'\tau){\cal F}({\bf r}\tau{\bf r}'\tau)=
{\bf S}_{h}({\bf r}\tau)\cdot{\bf S}_{h}({\bf r'}\tau )+\frac{1}{4}$, respectively.
The effective non-retarded interaction $J_A>0$ in front of the  
$\bar{\cal A}({\bf r}\tau{\bf r}'\tau){\cal A}({\bf r}\tau{\bf r}'\tau)$ term
constitutes the attractive potential for fermion pairing.
By HF decoupling of the  four-fermion
term in Eq.(\ref{pairing}) one obtains the ``$d$-wave" solution for the 
singlet pairing gap $\Delta_d({\bf r}{\bf r}')=\langle \bar{\cal A}({\bf r}\tau{\bf r}'\tau) \rangle$.
The role of antiferromagnetic correlations is also apparent due to the presence of spin-angular
correlation function in Eq.(\ref{jaf}). For example,
in a fully developed AF background 
${\bf\Omega}({\bf r}\tau)\cdot{\bf \Omega}({\bf r'}\tau )=-1$ for ${\bf r}$ and ${\bf r}'$
on neighboring sites, so that $J_A=4t^2/U$ and $J_F=0$, thus promoting  ``$d$-wave" pairing.
The expectation value of the second valence bond operator
 $\langle \bar{\cal F}({\bf r}\tau{\bf r}'\tau)\rangle$ competes with $\Delta_d({\bf r}{\bf r}')$, since it enhances the  kinetic energy of fermions, which eventually results in suppression of the $d$-wave gap at higher doping level.
The actual strength of the effective interaction in Eq.(\ref{jaf})
requires that the quantity $\langle{\bf\Omega}({\bf r}\tau)
\cdot{\bf \Omega}({\bf r'}\tau )  \rangle$ has to be  determined self-consistently.
To this end, in a way similar to Eq.(\ref{x1}), one should integrate from the effective Hamiltonian 
in Eq.(\ref{explicit})
charge and fermion variables to obtain the  action for the spin rotational degrees of freedom:
\begin{eqnarray}
 {\cal S}[{\bf \Omega}] &=&\frac{J}{4}
\sum_{\langle {\bf r}{\bf r}'\rangle}\int_0^\beta d\tau 
	\left[{\bf \Omega}({\bf r}\tau)\cdot{\bf \Omega}({\bf r'}\tau )-1\right]
	\nonumber\\
	&\equiv & J
\sum_{\langle {\bf r}{\bf r}'\rangle}\int_0^\beta d\tau 
	\left[{\bf S}_{\zeta}({\bf r}\tau)\cdot{\bf S}_{\zeta}({\bf r'}\tau )-\frac{1}{4}\right],
	\label{afaction}
\end{eqnarray}
where we made use of the formula
${\bf \Omega}=\frac{1}{2}{\rm tr}({\bf R}\hat{\sigma}^z{\bf R}^\dagger\hat{\bm\sigma})$,
 while  the variables ${\bf S}_{\zeta}({\bf r}\tau)=\frac{1}{2}\sum_{\alpha\gamma}\bar{\zeta}_{\alpha}({\bf r}\tau) 
\hat{\bm\sigma}_{\alpha\gamma} \zeta_{\gamma} ({\bf r}\tau)$ are the ``bosonic" spins
in the complex-projective (CP$^1$) formulation, see Eq.(\ref{cp1}). Here,
$J=\frac{4t^2}{U}(n_{h\uparrow}-n_{h\downarrow})^2$
indicates that Coulomb energy $U$ induced Hubbard band splitting is a necessary prerequisite
 to sustain AF correlations. Although, the fermions $\bar{h}_{{\alpha}}({\bf r}\tau),h_{\alpha}({\bf r}\tau)$ 
 play the role similar to ``spinons"
in the slave particle formulation a major quantitative difference appears. In SP theory at
half-filling ``spinons" are paired at $T=0$ with $\Delta_d\neq 0$. This is clearly
impossible here due to the presence  Zeeman-type band splitting term in Eq.(\ref{explicit})
that marks the onset of  charge gap.	
It prevents a non-zero solution for
$\Delta_d$ since at half filling $\rho$  is approaching the high energy charge gap,
 while $\Delta_d$ is governed by the exchange energy
$J_A\ll U$. We ascribe this discrepancy to the inherent inability in the  SP scheme
 to give an account of the high-energy effects that are the hallmark of the Mott physics.\cite{phillips} The reason is that the SP theory  handels exclusively with with a strictly {\it low-energy} effective Hamiltonian. This feature is  odd { e.g.} with 
experiments which  show that the superconducting transition in cuprates is accompanied by
changes  in the optical response, even at energies of the order of 100 times the critical temperature,
clearly pointing out the importance of the high-energy effects on the scale given by $U$.\cite{optical}
Finally we observe that superconductivity demands
more than just paired fermions - it also requires phase coherence in the charge sector
distinguished by the variables $z({\bf r}\tau)=e^{i\phi ({\bf r}\tau)}$.
Therefore a fully self-consistent theory requires a counterpart of the action in Eq.(\ref{afaction})
for the phase variables and phase stiffnesses that are responsible for the actual superconducting state.\cite{kopec} 

This work was supported by the Ministry of Education and Science
(MEN) under Grant No. 1 P03B 103 30  in the years 2006--2008.
%

\end{document}